\documentclass[11pt,canadian,preprint,preprintnumbers,nofootinbib,superscriptaddress]{revtex4}
\usepackage[latin9]{inputenc}
\usepackage[a4paper]{geometry}
\usepackage{color}
\usepackage{babel}
\usepackage{amsmath}
\usepackage{amssymb}
\usepackage{graphicx, subfigure,rotating}
\usepackage{esint}
\usepackage{epstopdf}

 \usepackage{wrapfig}

\makeatletter
\@ifundefined{textcolor}{}
{%
 \definecolor{BLACK}{gray}{0}
 \definecolor{WHITE}{gray}{1}
 \definecolor{RED}{rgb}{1,0,0}
 \definecolor{GREEN}{rgb}{0,1,0}
 \definecolor{BLUE}{rgb}{0,0,1}
 \definecolor{CYAN}{cmyk}{1,0,0,0}
 \definecolor{MAGENTA}{cmyk}{0,1,0,0}
 \definecolor{YELLOW}{cmyk}{0,0,1,0}
}

\usepackage{color}\usepackage{graphics}\usepackage{dcolumn}\usepackage{bm}\usepackage{afterpage}

\numberwithin{equation}{section}

\makeatother

\begin{document}
\allowdisplaybreaks

\title{Numerical Boson Stars with a Single Killing Vector II: the $D=3$ Case}

\vskip1cm
\author{Sean Stotyn}
\email{sstotyn@physics.ubc.ca}
\affiliation{Department of Physics and Astronomy, University of British Columbia,\\
                   Vancouver, British Columbia, Canada, V6T 1Z1}
\author{Melanie Chanona}
\email{mchanona@uwaterloo.ca}
\author{Robert B. Mann}
\email{rbmann@sciborg.uwaterloo.ca}
\affiliation{Department of Physics and Astronomy, University of Waterloo,\\
                   Waterloo, Ontario, Canada, N2L 3G1}

\begin{abstract}

We complete the analysis of part I in this series (Ref. \cite{Stotyn:2013yka}) by numerically constructing boson stars in 2+1 dimensional Einstein gravity with negative cosmological constant, minimally coupled to a complex scalar field.  These lower dimensional boson stars have strikingly different properties than their higher dimensional counterparts, most noticeably that there exists a finite central energy density, above which an extremal BTZ black hole forms.  In this limit, all of the scalar field becomes enclosed by the horizon; it does not contract to a singularity, but rather the origin remains smooth and regular and the solution represents a spinning boson star trapped inside a degenerate horizon. Additionally, whereas in higher dimensions the mass, angular momentum, and angular velocity all display damped harmonic oscillations as functions of the central energy density, in $D=3$ these quantities change monotonically up to the bound on the central energy density.  Some implications for the holographic dual of these objects are discussed and it is argued that the boson star and extremal BTZ black hole phases are dual to a spontaneous symmetry breaking at zero temperature but finite energy scale.

\end{abstract}

\maketitle

\section{Introduction}

Boson stars, \emph{i.e.} smooth horizonless geometries composed of localized bundles of self-interacting scalar fields, have attracted attention over the years both from the viewpoint of astrophysics (for instance as candidates for gravitationally bound dark matter) and from their connection to interesting phases of gauge theories through the anti-de Sitter/conformal field theory (AdS/CFT) correspondence -- see Ref. \cite{Liebling:2012fv} for a general review.  In 3 dimensions, exact non-rotating boson star solutions with negative cosmological constant that are comprised of a massive complex scalar field with a quartic self-interaction have been known for some time, although only in the limit of large self-coupling\cite{Sakamoto:1998hq,Sakamoto:1998aj}.  Subsequently, rotating boson stars composed of a massive complex scalar field with arbitrary quartic self-coupling were numerically studied in \cite{Astefanesei:2003rw} and the authors found explicitly that solutions only exist for a non-vanishing negative cosmological constant.  Consequently, boson star solutions in 3 dimensions are only known to exist in asymptotically AdS spaces and can thus provide interesting information about conformal field theories in 2 dimensions.  

In their analysis, the authors of \cite{Astefanesei:2003rw} made all physical quantities dimensionless via normalizing by the scalar field mass, making the limit of vanishing scalar field mass intractable.  In the present paper we remedy this shortcoming by explicitly constructing asymptotically AdS boson star solutions in 3 dimensions using a massless complex scalar field.  We furthermore use the AdS length as the parameter by which we normalize all of the physical quantities.  In so doing, we find qualitative agreement with the results of \cite{Astefanesei:2003rw} but we further discover striking new features that were previously hidden.  For instance, just as for a massive scalar field we find that for a massless scalar field, the boson star mass, angular momentum, and angular velocity are all monotonic functions of the central energy density, unlike the cases with large self interaction \cite{Sakamoto:1998hq} and in higher dimensions \cite{Stotyn:2013yka,Dias:2011at}, where all of these quantities exhibit damped harmonic oscillations about finite central values.  A striking new feature that appears in our analysis, however, is that there is an upper bound on the central energy density, at which a degenerate horizon forms around the boson star: that is, it becomes an extremal black hole.  

Black hole solutions in 3 dimensions are considerably different than in $D>3$. Most notably that they can only exist for a non-vanishing negative cosmological constant.  First discovered by Ba\~nados, Teitelboim, and Zanelli (BTZ) in \cite{Banados:1992wn}, the BTZ black hole shares many properties in common with its higher dimensional counterparts, yet it possesses key features that set it apart.  For one, it is a space of constant negative curvature and is hence everywhere locally a patch of AdS space.  Secondly, the zero mass BTZ black hole is disconnected from the AdS vacuum by a mass gap; vacuum states within the mass gap correspond to naked conical singularities, although we will show that the mass gap is also continuously populated by a family of boson star solutions that do not contain a conical singularity.
These two properties stem from the fact that the BTZ black hole is obtained by identifying points in global AdS$_3$ related by the action of a subgroup of the conformal group: a set of these identifications persist even in the limit of vanishing black hole mass parameter.

The rotating, non-extremal BTZ black hole solution is given explicitly by
\begin{equation}
ds^2=-f(r)dt^2+\frac{dr^2}{f(r)}+r^2\left(d\phi-\Omega(r)dt\right)^2
\end{equation}
\begin{equation}
f(r)=\frac{r^2}{\ell^2}-M+\frac{J^2}{4r^2}, \qquad\qquad \Omega(r)=\frac{J}{2r^2} \label{eq:BTZmetric}
\end{equation}
where $M$ and $J$ are the ADM mass and angular momentum respectively.  There is no curvature singularity present at $r=0$, but rather a singularity in the group action as described above.  It was further shown in \cite{Ross:1992ba} that the BTZ black hole can form from a collapsing shell of dust.  In the latter case, the singularity in the group action at $r=0$ is replaced by a delta function source whose physical origin is the collapsed matter.  

This illustrates an important point: the structure inside the horizon depends on the analytic continuation of the exterior solution.  Furthermore, this analytic continuation is not necessarily unique in general.  Some extra physical input is needed to determine the internal structure -- in the above example, this corresponds to whether the black hole is eternal or is formed from collapsing matter.  In the present paper, we obtain numerical evidence for another concrete realization of this ambiguity through the limit in which the boson star collapses to an extremal black hole.  Outside the horizon the geometry is identical to extremal BTZ, while inside the horizon there is an everywhere regular solution with a nontrivial scalar field whose stress energy sharply decays as the horizon is approached.  The presence of the scalar field changes the internal geometry such that the singularity at the origin is removed.

The rest of the paper is organized as follows: in section \ref{sec:setup} we discuss the physical setup and the corresponding equations of motion.  In section \ref{sec:boundaryconditions} we discuss the boson star boundary conditions imposed at the origin and asymptotically, as well as the asymptotic charges entering the first law of thermodynamics.  Next we discuss the set up for numerically solving the equations of motion in section \ref{sec:results} and present our numerical results.  Finally in section \ref{sec:discussion} we discuss the implications of our results and conclude with some open questions and future work needed.

\section{Setup}
\label{sec:setup}

We begin with 3 dimensional Einstein gravity with negative cosmological constant, $\Lambda=-\frac{1}{\ell^2}$, minimally coupled to a complex scalar field
\begin{equation}
S=\frac{1}{16\pi}\int{d^3x\sqrt{-g}\left(R+\frac{2}{\ell^2}-2\big|\nabla\Pi\big|^2\right)}.\label{eq:action}
\end{equation}
Following the 3D perturbative analysis set out in \cite{Stotyn:2012ap}, we consider the metric and scalar field ans\"atze
\begin{equation}
ds^2=-f(r)g(r)dt^2+\frac{dr^2}{f(r)}+r^2\big(d\phi-\Omega(r) dt\big)^2\label{eq:metric}
\end{equation}
\begin{equation}
\Pi=\Pi(r) e^{-i\omega t+i\phi} \label{eq:ScalarField}
\end{equation}
In these coordinates, $\phi$ has range $[0,2\pi]$ and the radial coordinate has range $r\in[0,\infty)$.  This form of the metric and scalar field was also considered in \cite{Astefanesei:2003rw} for the action (\ref{eq:action}) with an additional potential term for the scalar field, $V(\Pi)$.  The authors found rotating boson star solutions by numerically solving the corresponding equations of motion explicitly for $V(\Pi)=\mu^2 |\Pi|^2$, although they claim to have also solved the equations of motion for more complicated forms of $V$.  For the quadratic potential, all of the physical quantities were made dimensionless via a scaling by the mass parameter $\mu$.  Therefore, a straightforward limit of $\mu\rightarrow 0$ cannot be taken from the results of \cite{Astefanesei:2003rw}; the current paper addresses this directly by explicitly considering a massless scalar field.

The matter stress tensor is given by
\begin{equation}
T_{ab}=\left(\partial_a\Pi^*\partial_b\Pi+\partial_a\Pi\partial_b\Pi^*\right)-g_{ab}\left(\partial_c\Pi\partial^c\Pi^*\right),\label{eq:Tab}
\end{equation}
which has the same symmetries as the metric (\ref{eq:metric}), although the scalar field does not: the metric (\ref{eq:metric}) is invariant under $\partial_t$, $\partial_\phi$ while the scalar field (\ref{eq:ScalarField}) is only invariant under the combination
\begin{equation}
K=\partial_t+\omega\partial_\phi. \label{eq:KV}
\end{equation}
Therefore, any solution with nontrivial scalar field will only be invariant under the single helical Killing vector field given by (\ref{eq:KV}).  

The equations of motion resulting from the action (\ref{eq:action}) are $G_{ab}-\frac{1}{\ell^2}g_{ab}=T_{ab}$ and $\nabla^2\Pi=0$, which yield a system of coupled second order ordinary differential equations (ODEs)
\begin{equation}
f''+f'\left(\frac{3}{r}-\frac{g'}{2g}\right)+\frac{fg'}{g}\left(\frac{1}{r}-\frac{g'}{2g}\right)+\frac{8\Pi'\Pi}{r}+\frac{4\Pi^2}{fg}(\omega-\Omega)^2+\frac{4\Pi^2}{r^2}-\frac{8\Pi^2\Omega'r}{fg}(\omega-\Omega)-\frac{8}{\ell^2}=0, \label{eq:fEq}
\end{equation}
\begin{equation}
g''+g'\left(\frac{2f'}{f}+\frac{1}{r}\right)-\frac{8\Pi^2}{f^2}(\omega-\Omega)^2+\frac{8r\Pi^2\Omega'}{f^2}(\omega-\Omega)-\frac{8g\Pi'\Pi}{rf}=0, \label{eq:gEq}
\end{equation}
\begin{equation}
\Omega''+\frac{4\Pi^2}{fr^2}(\omega-\Omega)+\Omega'\left(\frac{3}{r}-\frac{g'}{2g}\right)=0, \label{eq:OmegaEq}
\end{equation}
\begin{equation}
\Pi''+\frac{\Pi'(f^2gr^2)'}{2f^2gr^2}+\frac{\Pi}{f^2g}(\omega-\Omega)^2-\frac{\Pi}{fr^2}=0, \label{eq:PiEq}
\end{equation}
where a $'$ denotes differentiation with respect to $r$.  In addition to these second order ODEs, the Einstein equations further impose two first order ODEs in the form of constraint equations, $C_1=0$ and $C_2=0$. Explicitly, these are
\begin{equation}
C_1=\frac{r}{f}\left(f^2g\right)'+4g\left(\Pi^2-\frac{r^2}{\ell^2}\right)+r^4\Omega'^2, \label{eq:C1Eq}
\end{equation}
\begin{equation}
C_2=\frac{\Pi^2(\omega-\Omega)^2}{f^2g}+\Pi'^2+\frac{r^2\Omega'^2}{4fg}+\frac{f'}{2fr}+\frac{\Pi^2}{fr^2}-\frac{1}{f\ell^2}. \label{eq:C2Eq}
\end{equation}
Under the flow of the equations of motion, the constraint equations satisfy the relations,
\begin{equation}
C_1'={\cal F}_1(r)C_1,
\end{equation}
\begin{equation}
C_2'={\cal F}_2(r)C_1+{\cal F}_3(r)C_2,
\end{equation}
for some functions ${\cal F}_1(r),~{\cal F}_2(r),~\mathrm{and}~{\cal F}_3(r)$.  The explicit form of these functions is largely unimportant: the fact that the derivatives of the constraints are proportional to the constraints themselves implies that satisfying the constraints at some point ensures they are satisfied everywhere.  In practice we impose the constraint equations at the boundary points, guaranteeing that they continue to hold everywhere.

\section{Boundary Conditions}
\label{sec:boundaryconditions}

In order to solve the equations of motion (\ref{eq:fEq})-(\ref{eq:PiEq}) to obtain boson star solutions, we need to impose appropriate boundary conditions at the boson star origin and asymptotically.  In this section, we detail these boundary conditions, starting with the origin. 

\subsection{Boundary Conditions at the Origin}

Boson stars are smooth, horizonless geometries, which means that all metric functions must be regular at the origin.  To find the boundary condition on $\Pi$, we multiply (\ref{eq:PiEq}) by $r^2$ and note that $\Pi$ must vanish at the origin in order to yield consistent equations of motion.   Thus, the boundary conditions at the boson star origin take the form
\begin{eqnarray}
\left.f\right|_{r \rightarrow 0} = 1 + \mathcal{O}(r^2), \quad 
\left.g\right|_{r \rightarrow 0} = g(0)+\mathcal{O}(r^2), \quad 
\left.\Omega\right|_{r \rightarrow 0} = \Omega(0)+\mathcal{O}(r^2), \quad 
\left.\Pi\right|_{r \rightarrow 0} = q_0\frac{r}{\ell}, \label{eq:OriginBC}
\end{eqnarray}
where $q_{0}$ is a dimensionless parameter
such that the energy density of the scalar field, $T^{00}$, at the
origin is proportional to $q_{0}^{2}$. In fact, $q_{0}$ uniquely
parameterizes the one-parameter family of boson star solutions in
each dimension.  Formally, it is defined by $q_0\equiv\ell\Pi'(0)$.

\subsection{Asymptotic Boundary Conditions}

In order to simplify the asymptotic boundary conditions, we first make note of a residual gauge freedom.  It is straightforward to show that the transformation
\begin{eqnarray} \label{PsiGaugeFreedom}
\phi = \tilde\phi + \lambda t, \quad\quad \Omega = \tilde\Omega + \lambda, \quad\quad \omega = \tilde\omega + \lambda
\end{eqnarray}
for some arbitrary constant $\lambda$, leaves both the metric (\ref{eq:metric}) and scalar field (\ref{eq:ScalarField}) unchanged.  We find it convenient in our numerical analysis to set $\lambda=\omega$ so that we can set $\tilde{\omega}=0$: in this frame, the coordinates are rigidly
rotating asymptotically so that $\tilde{\Omega}(r)\rightarrow-\omega$ as $r\rightarrow\infty$.  In what follows, we use $\tilde\phi$, $\tilde \Omega(r)$, and $\omega$ unless otherwise stated, however we drop the tildes for notational convenience.

In this limit the boundary conditions will be the same for the boson star and the black hole, in particular they will asymptote to the BTZ metric (\ref{eq:BTZmetric}) with higher order multipole corrections:
\begin{align} 
\left.f\right|_{r \rightarrow \infty}= \frac{r^2}{\ell^2} -M  + \mathcal{O}(r^{-2}), \quad\>\> 
\left.g\right|_{r \rightarrow \infty} &= 1  + \mathcal{O}(r^{-4}), \quad\>\>
\left.\Omega\right|_{r \rightarrow \infty} = -\omega+\frac{J}{2r^2} + \mathcal{O}(r^{-4}), \label{eq:AsymptoticBC}\\ 
\left.\Pi\right|_{r \rightarrow \infty}&= \frac{\epsilon \ell^2}{r^2} + \mathcal{O}(r^{-3}),\nonumber
\end{align}
where $M$ and $J$ are the ADM mass and angular momentum of the full solution respectively.  The boundary condition on $\Pi$ is set by requiring it to be normalizable.  Here, $\epsilon$ is a dimensionless measure of the amplitude of the scalar field at infinity. In the perturbative analysis of \cite{Stotyn:2011ns}, $\epsilon$ uniquely parameterizes the boson star
solutions but in the non-perturbative regime it does not, as will be explicitly demonstrated in the following section.

Finally, note that the scalar field vanishes asymptotically, implying the existence of two asymptotic Killing vectors, $\partial_t$ and $\partial_\phi$.  Indeed, ${M}$ and ${J}$ are the conserved charges associated with the asymptotic symmetries $\partial_t$ and $\partial_\phi$ respectively.  These asymptotic Killing symmetries imply that the boson stars obey the first law of thermodynamics
\begin{equation}
d{M}=\omega d{J},
\end{equation}
where the $TdS$ term is missing because these are horizonless objects.

\section{Results}
\label{sec:results}

The procedure we used to solve the equations of motion are identical to the higher dimensional case in \cite{Stotyn:2013yka}, namely a relaxation method on a Chebyshev grid.  The numerical tests for solution validity are identical as well, so instead of repeating it here we refer the reader to Ref. \cite{Stotyn:2013yka} for the details.  For completeness, in the first subsection we outline the details in the numerical setup that differ from the higher dimensional case.  Then we present our results for boson stars in $D=3$.

\subsection{Numerical Details Specific to $D=3$}

To use the Chebyshev relaxation procedure, we need to compactify the domain of integration, which is done by introducing the coordinate, $y\in[0,1]$:
\begin{equation}
y=\frac{r^2}{r^2+\ell^2}.
\end{equation}
In terms of this coordinate, the Chebyshev grid is defined by the set of $N+1$ points $y_k=\frac{1+x_k}{2}$, $k=0,...,N$, where $x_k=\cos\left(\frac{k\pi}{N}\right)$ are the extrema of of the $N^{\mathrm{th}}$ Chebyshev polynomial $T_N(x)$.

To obtain a set of analytic functions, we first extract the singular
behavior from each and introduce auxiliary functions with the boundary
conditions (\ref{eq:OriginBC}) and (\ref{eq:AsymptoticBC}) in mind. In terms of the coordinate $y$, this leads to
\begin{align}
&f(y)=\frac{y}{1-y}+q_f(y),\\
&g(y)=1+(1-y)q_g(y),\\
&\Omega(y)=\frac{q_\Omega(y)}{\ell},\\
&\Pi(y)=\sqrt{y}(1-y)q_\Pi(y).
\end{align}
The set of functions $\{q_{f},q_{g},q_{\Omega},q_{\Pi}\}$ are analytic over the range $y\in[0,1]$; in the remaining discussion the coordinate $y$ will be used, unless otherwise stated. In particular,
a prime $'$ will denote a derivative with respect to $y$.

To find the boundary conditions on the $q$ functions, we Taylor expand the equations of motion (\ref{eq:fEq})--(\ref{eq:PiEq}) and the constraint equations (\ref{eq:C1Eq}) and (\ref{eq:C2Eq}) around the two boundary points, $y_0=0,1$ and require them to vanish order by order in $(y-y_0)$. This leads to the following nontrivial relationships between the various functions and their first derivatives: at $y_0=0$ 
\begin{align}
&q_f(0)-1=0, \qquad\qquad q_g'(0)-q_g(0)-2\big(1+q_g(0)\big)q_0^2=0,\\
&q_\Pi(0)-q_0=0, \qquad\qquad q_\Omega'(0)-\frac{1}{2}q_0^2q_\Omega(0)=0,
\end{align}
and at $y_0=1$ 
\begin{align}
&q_f'(1)+q_\Omega'(1)^2+4q_\Pi(1)^2=0, \qquad\qquad q_g'(1)-4q_\Pi(1)^2=0,\\
&q_g(1)=0, \qquad\qquad q_\Pi'(1)+\frac{1}{8}\big(13-4q_f(1)-q_\Omega(1)^2\big)q_\Pi(1)=0.
\end{align}

With these specific details used in the numerical construction presented in \cite{Stotyn:2013yka}, we can now present our results for $D=3$.

\subsection{Numerical Results}

We note that the physical properties of the boson stars can be written in terms of the boundary values of the analytic $q$ functions.  To make all physical quantities dimensionless, we work in units where $\ell=1$.  The Kretschmann invariant, $K=R_{abcd}R^{abcd}$ at the centre of the boson star is given by
\begin{equation}
K=12-16q_0^2+16q_0^4,
\end{equation}
while the mass, angular momentum, angular velocity, and perturbative parameter $\epsilon$ used in \cite{Stotyn:2012ap} are given by
\begin{equation}
{M}=-q_f(1), \qquad\qquad {J}=-2 q_\Omega'(1), \qquad\qquad \omega=-q_\Omega(1), \qquad\qquad \epsilon=q_\Pi(1).
\end{equation}
\begin{figure}[ht!]
     \begin{center}
        \subfigure[The boson star mass against the perturbative parameter $\epsilon$.  Empty AdS$_3$ is given by $M=-1$.]{%
            \label{fig:first}
            \includegraphics[width=0.475\textwidth]{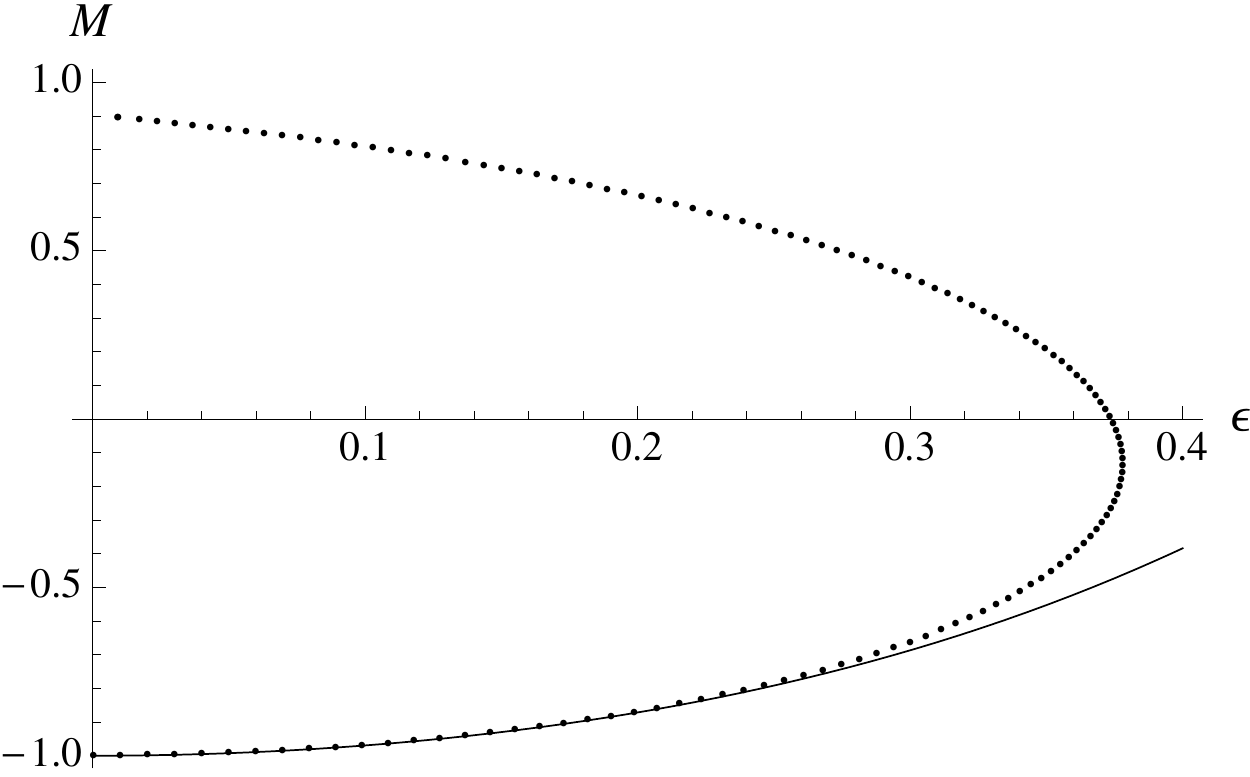}
        }%
        \subfigure[The boson star angular momentum against the perturbative parameter $\epsilon$.]{%
           \label{fig:second}
           \includegraphics[width=0.475\textwidth]{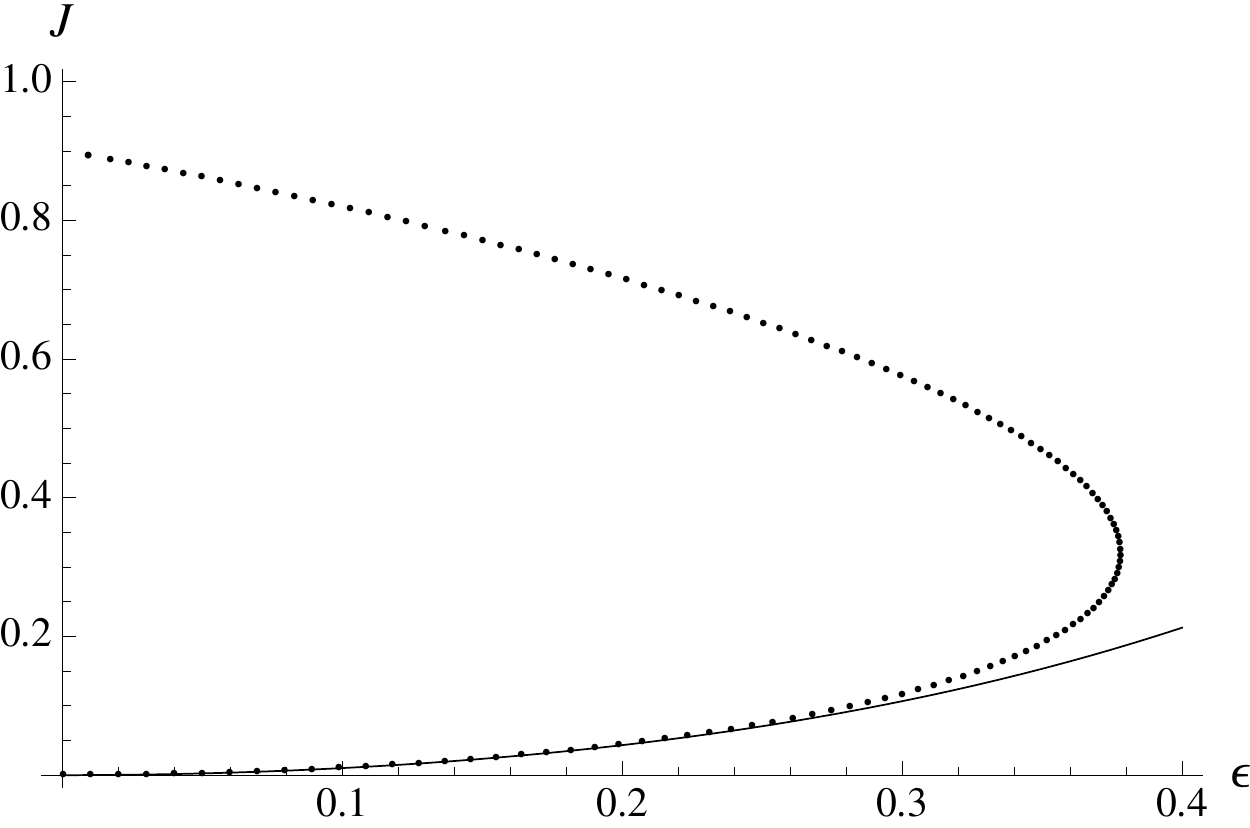}
        }\\ %  ------- End of the first row ----------------------%
        \subfigure[The boson star angular velocity against the perturbative parameter $\epsilon$.]{%
            \label{fig:third}
            \includegraphics[width=0.475\textwidth]{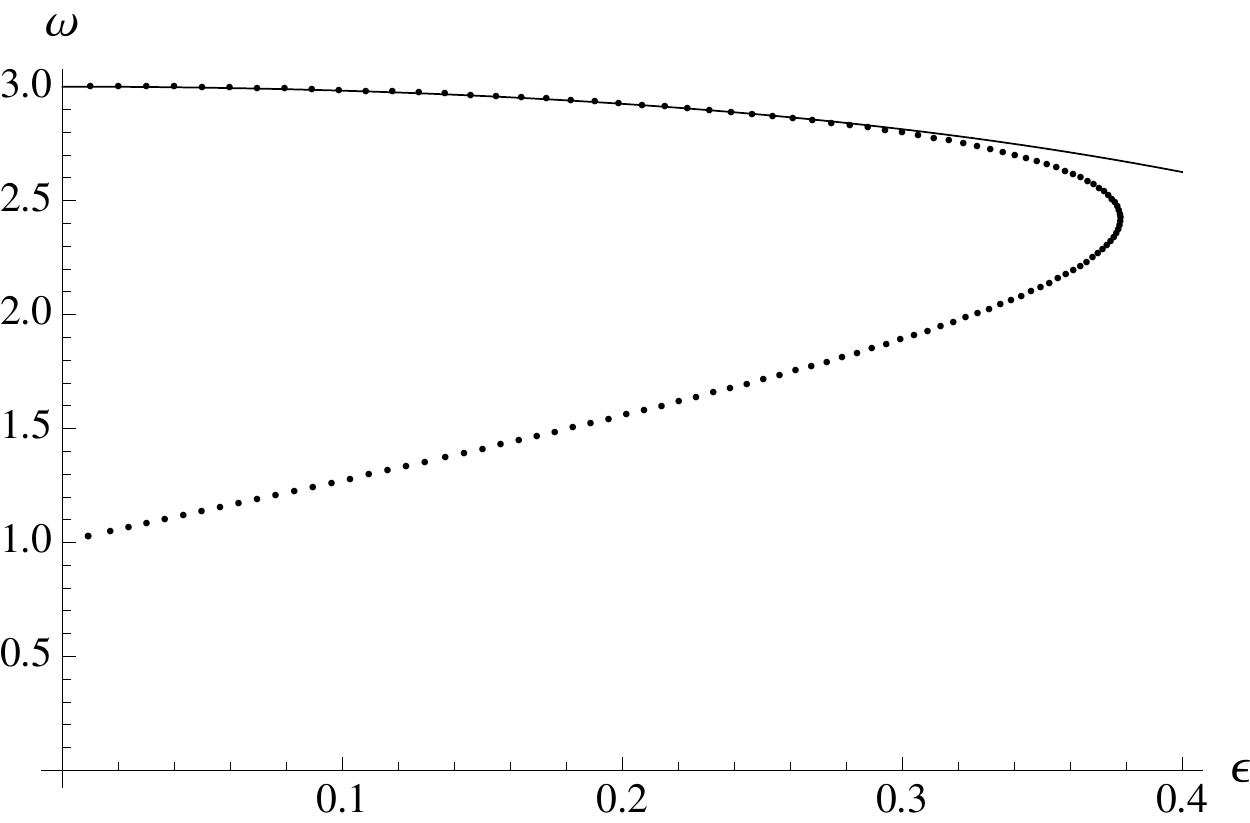}
        }%
        \subfigure[The Kretschmann scalar at the centre of the boson star plotted against the central energy density.]{%
            \label{fig:fourth}
            \includegraphics[width=0.475\textwidth]{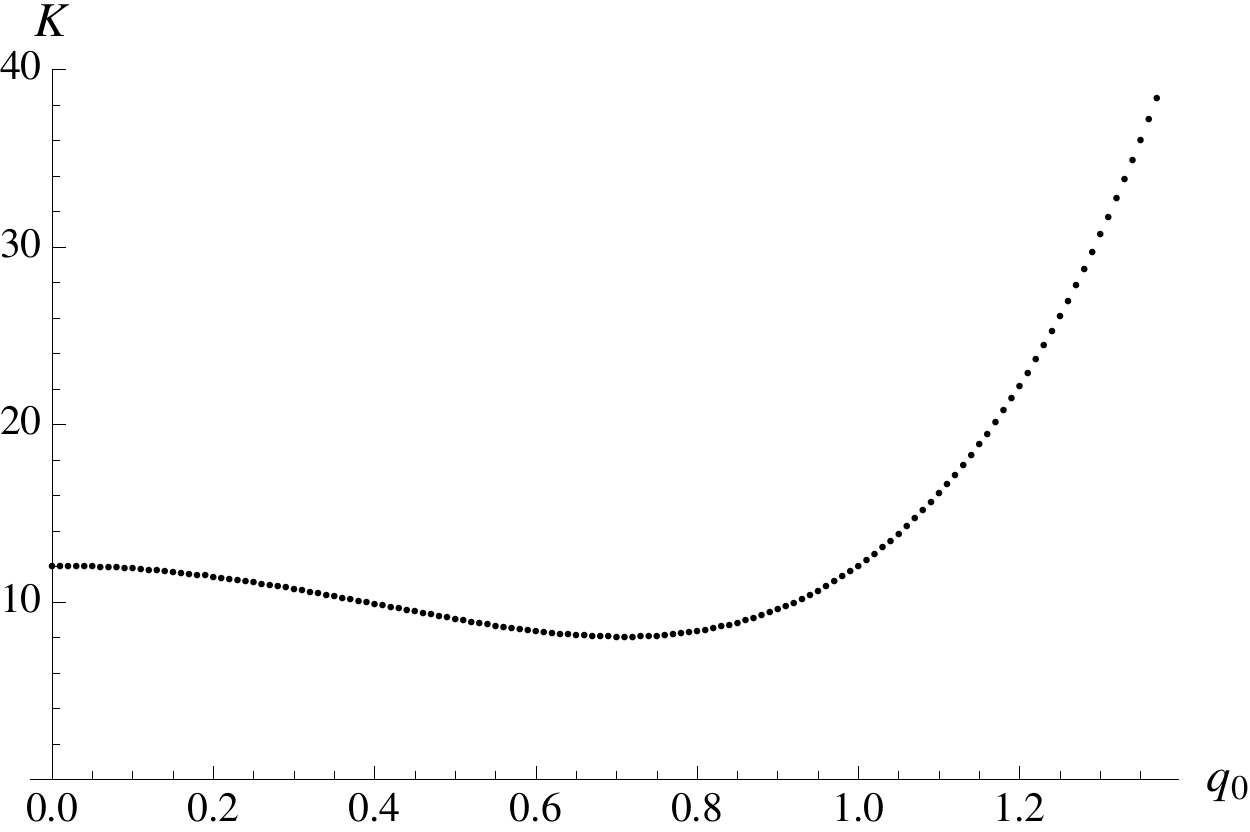}
        }%
    \end{center}
    \caption{%
        The physical properties of the 3D boson stars up to the limiting central energy density.  Each dot represents a numerical solution at a particular value of $q_0$ while the solid lines are the perturbative results of \cite{Stotyn:2012ap}.
     }%
   \label{fig:plots1}
\end{figure}
These physical quantities are plotted in figure \ref{fig:plots1} for the whole range of allowed boson star solutions, along with the perturbative results of \cite{Stotyn:2012ap} (solid lines).
Figures \ref{fig:first}, \ref{fig:second}, and \ref{fig:third} show the mass, angular momentum, and angular velocity plotted against the asymptotic scalar field strength.  Empty AdS$_3$ corresponds to $M=-1$, $J=0$, and $\omega=3$, while each subsequent data point is at a higher value of $q_0$ than the previous with a step size of $\Delta q_0=0.01$.  The first feature of note is that for boson stars in $D=3$, these physical quantities are monotonic in $q_0$.  This is in contrast to higher dimensional boson stars, which exhibit damped harmonic oscillations about finite limiting values\cite{Stotyn:2013yka,Dias:2011at}.  The maxima of these oscillations in parameter space for higher dimensional boson stars were conjectured to be unstable to the formation of a black hole.  However, in 3 dimensions such a potential instability does not arise because of the monotonic increase of the mass with the central energy density.  This is in general agreement with the results of \cite{Stotyn:2012ap} where it was found that perturbative hairy black holes do not exist in 3 dimensions.

\begin{figure}[ht!]
     \begin{center}
	\subfigure[~The scalar field function $\Pi(r)$ plotted for the values of $q_0$ indicated.]{%
            \label{fig:third2}
            \includegraphics[width=0.475\textwidth]{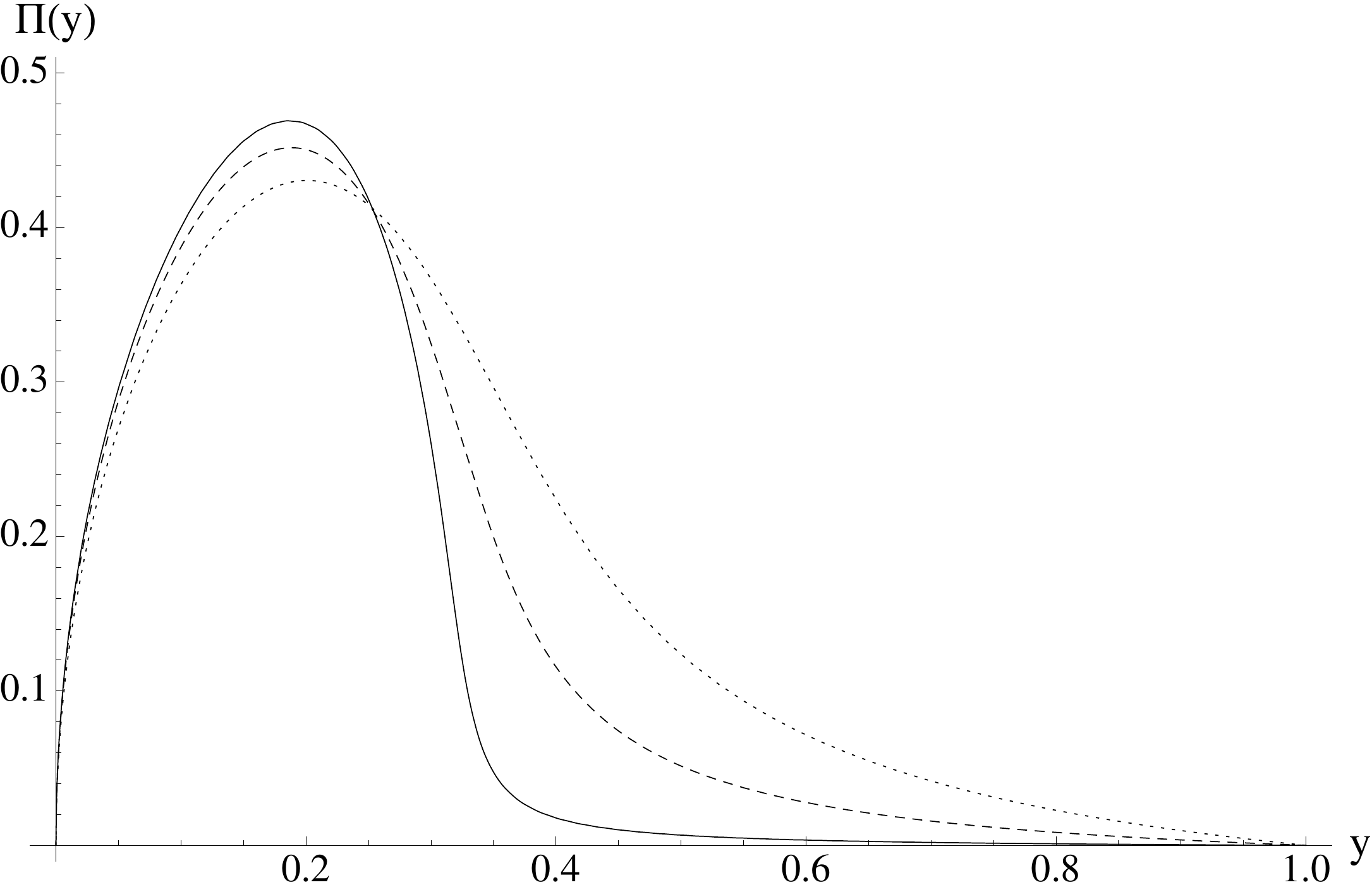}
            \put(-128,57){\begin{turn}{-45} \tiny $q_0=1.25$ \end{turn}}
            \put(-133,35){\begin{turn}{-30} \tiny $q_0=1.33$\end{turn}}
            \put(-142,17){\begin{turn}{-10} \tiny $q_0=1.38$ \end{turn}}
        }%
        \subfigure[~$M/J$ plotted against $q_0$ approaching unity.]{%
            \label{fig:first2}
            \includegraphics[width=0.475\textwidth]{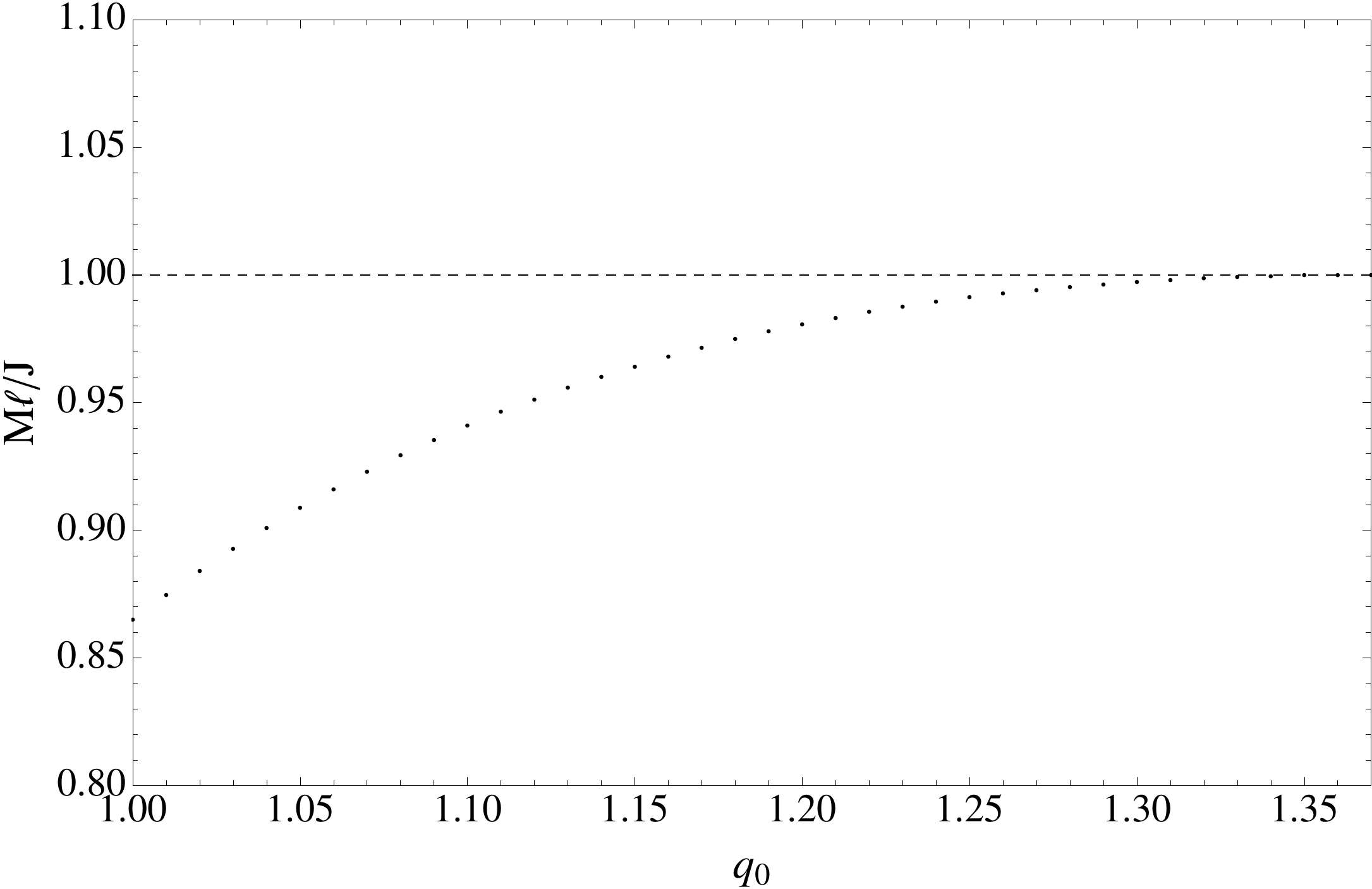}
        }\\ %  ------- End of the first row ----------------------%
        \subfigure[~The metric function $f(r)$ plotted for the values of $q_0$ indicated.]{%
           \label{fig:second2}
           \includegraphics[width=0.45\textwidth]{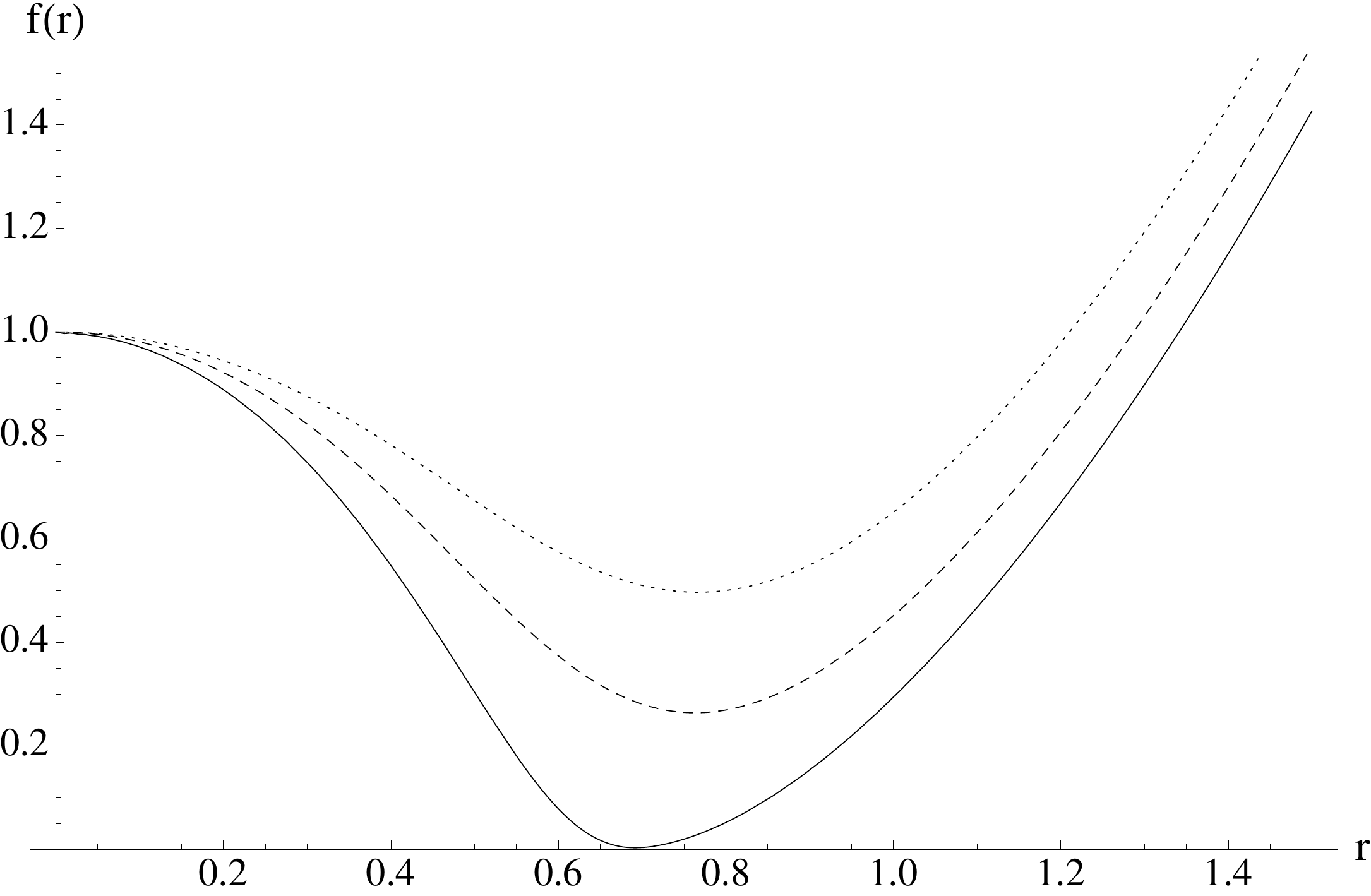}
           \put(-123,55){\tiny $q_0=1.09$}
           \put(-123,36){\tiny $q_0=1.21$}
           \put(-130,16){\tiny $q_0=1.38$}
         }%
         \subfigure[~The metric function $f(r)$ for $q_0=1.38$ (solid) and for an extremal BTZ black hole with $M=0.91$ (dashed).]{%
            \label{fig:fourth2}
            \includegraphics[width=0.475\textwidth]{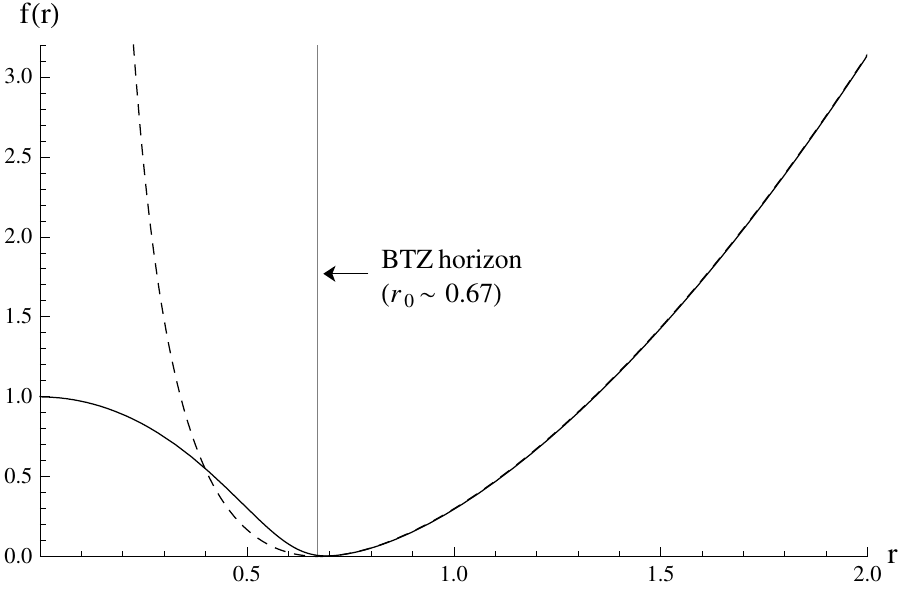}
        }%
    \end{center}
    \caption{%
        Four independent indications that the boson star is limiting to an extremal BTZ black hole as the maximum value of $q_0$ is approached: $(a)$ the scalar field is piling up behind the soon-to-be horizon, $(b)$ the solution is approaching $M=J$, $(c)$ the metric function $f(r)$ is forming a degenerate zero at a finite $r$,  and $(d)$ the metric function $f(r)$ is approaching the extremal BTZ black hole function exterior to the degenerate horizon.
     }%
   \label{fig:plots2}
\end{figure}

The second noteworthy feature in 3 dimensions is that there is a maximum value of $q_0\approx 1.38$, beyond which these boson stars do not exist; this upper bound on the central energy density is not present in higher dimensions.  The 3D solution is limiting to one for which $M,~J,$ and $\omega$ all have finite limiting values but the asymptotic strength of the scalar field vanishes.  By performing a power series expansion of the equations of motion at the boundary $y=1$, one can show that if $\epsilon$ vanishes asymptotically, then in this limit the scalar field must vanish everywhere on the domain of integration by analyticity.  As can be seen in figure \ref{fig:third2}, as the critical value of $q_0$ is reached, the scalar field becomes concentrated into an increasingly smaller finite region, outside of which it vanishes.  We now show that in this limit, the boson star becomes enveloped by a degenerate horizon and so collapses to an extremal black hole.  For such an extremal black hole, the domain of integration is $r\in[r_+,\infty)$, so by analyticity and $\epsilon$ vanishing in the limit, the scalar field must vanish everywhere outside the horizon, \emph{i.e.} the black hole cannot have scalar hair.

For an extremal BTZ black hole, the parameters must satisfy $M=J$ and $\omega=1$, where $\omega$ is the horizon angular velocity.  As can be seen in figure \ref{fig:third}, as the limiting value of $q_0$ is approached, $\omega$ linearly approaches unity.  Similarly, figure \ref{fig:first2} shows that $M/J$ also approaches unity in the limit.  In figure \ref{fig:second2} we plot the full boson star metric function $f(r)$, showing that it forms a degenerate root at a finite value of $r$ as $q_0$ increases to its critical value.  From these lines of evidence, we conclude that as $q_0$ increases to its maximum, the scalar field condenses down into a smaller and smaller region until  it becomes trapped behind a degenerate horizon, leaving no remaining scalar field outside.

When this happens, the solution exterior to the horizon becomes identical to the extremal BTZ solution, while the solution interior to the horizon differs significantly.  This can be seen in figure \ref{fig:fourth2} in which we plot the boson star metric function $f(r)$ at $q_0=1.38$ (solid) along with the extremal BTZ metric function for $M=0.91$ (dashed).  The two metric functions agree in the region exterior to the degenerate root, although the interior structure is markedly different.  One shortcoming of the present work is that our numerical methods are ill-equipped to fully resolve this transition from boson star to extremal black hole because the boundary conditions for both objects are completely different, as is the domain of integration.  Nevertheless, the evidence is strong that such a transition indeed occurs at finite central energy density.  Furthermore, in this transition the Kretschmann scalar at the boson star origin remains finite, as can be seen in figure \ref{fig:fourth}, as opposed to higher dimensions where a curvature singularity forms \cite{Stotyn:2013yka}.  One might have naively expected a delta function singularity to form at $r=0$ in the extremal limit, analogous to what happens with the collapsing dust considered in \cite{Ross:1992ba}.  However, it appears that this is not the case, and rather the origin remains a regular point of the interior solution, leading to the interpretation that the extremal BTZ black hole that forms can be thought of as a boson star trapped behind a degenerate horizon.

 \section{Discussion}
 \label{sec:discussion}
 
 We have completed the analysis started in \cite{Stotyn:2013yka} by numerically constructing single Killing vector boson stars in 3 dimensions.  Although the zero mass BTZ black hole is separated from pure AdS by a mass gap, the zero mass boson star is not.  We have shown that the entire mass gap can be populated by boson stars, whose stress energy is sufficient to smooth out the conical singularity normally present at $r=0$.  These boson stars are composed of a self-gravitating massless complex scalar field with harmonic time dependence.  They differ rather significantly from their higher dimensional counterparts found in \cite{Stotyn:2013yka,Dias:2011at}, most notably in the existence of a critical central energy density at which an extremal BTZ black hole forms.  The mass, angular momentum, and angular velocity all change monotonically with the central energy density up to this critical value, whereas in higher dimensions the corresponding quantities all exhibit damped harmonic oscillations about finite limiting values.  These 3D properties are all qualitatively shared by the 3D boson star solutions found in \cite{Astefanesei:2003rw}, which are composed of a massive complex scalar field.  Quantitatively the two are distinct as the scalar field mass causes considerably different fall-off behavior for the scalar field.  Nevertheless, it appears that the limit of vanishing scalar field mass is a smooth limit and reduces to our solution presented here, as it should.

\begin{wrapfigure}{l}{0.5 \textwidth}
\vspace{-0.5cm}
\center{\includegraphics[width=0.5\textwidth]{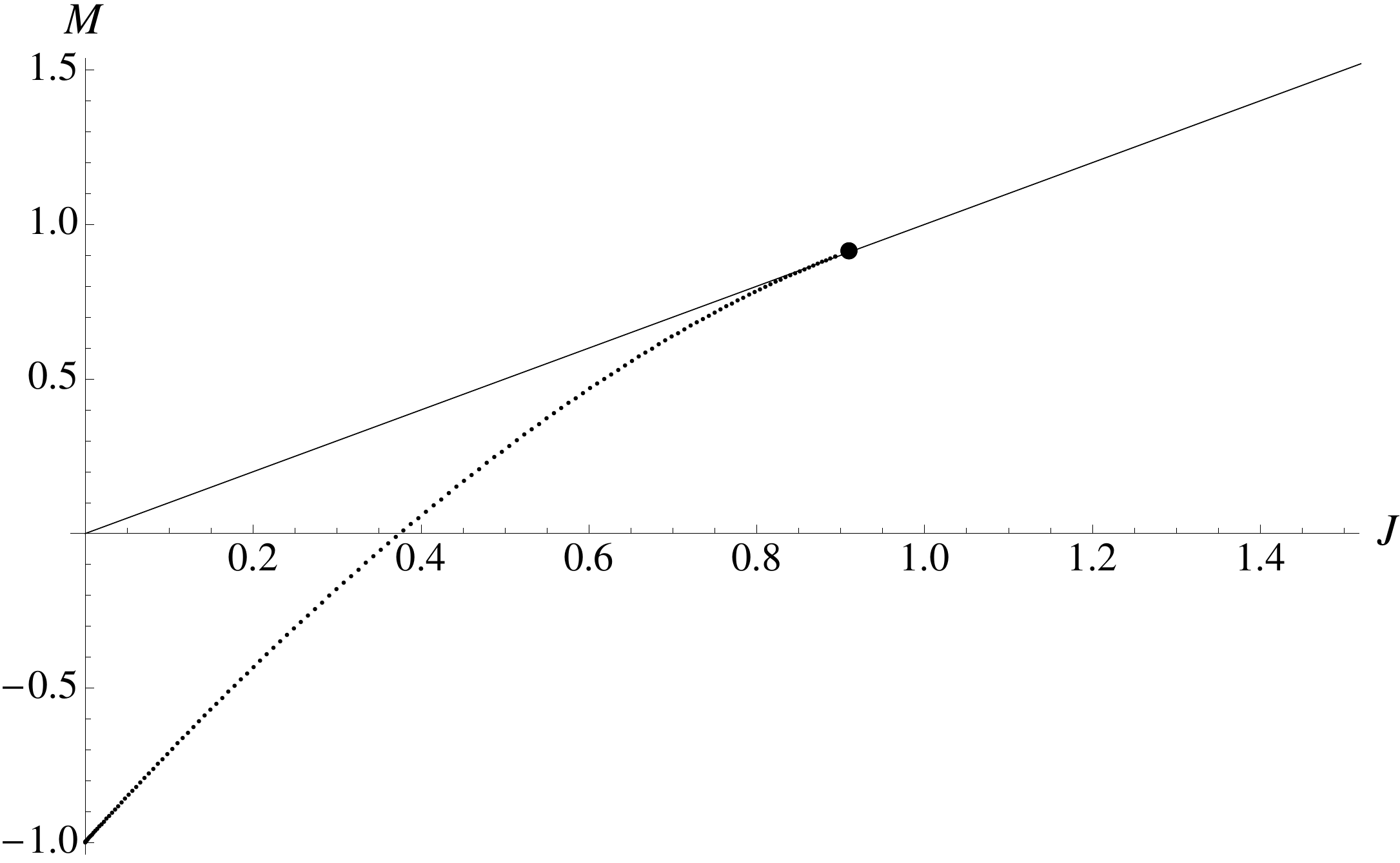}}
\caption{The phase diagram at zero temperature with extremal BTZ black holes (solid line) and boson stars (dotted line).  The large dot is the critical point where the boson star line ends.}
\label{fig:phase}
\end{wrapfigure}
Arguably the most interesting result of this paper is the formation of a degenerate horizon around the boson star at the critical point.   The extremal black hole, being devoid of scalar hair, is invariant under both the holomorphic and anti-holomorphic Killing fields, $K_1=\partial_t+\omega\partial_\phi$ and $K_2=\partial_t-\omega\partial_\phi$, whereas the boson stars are only invariant under the holomorphic Killing field $K_1$.  Furthermore, the boson stars only exist below the critical energy scale $M_0\approx0.91$.  The phase diagram of boson stars and extremal BTZ black holes is shown in figure \ref{fig:phase}: since the free energy is given by ${\cal F}=M-TS$ and both objects are at zero temperature, the vertical axis is the free energy and it is clear that the boson star is thermodynamically favored over the extremal black hole.  Above the critical point the scalar field vanishes and the full symmetry is present, whereas below the critical point the symmetry is reduced and a non-trivial scalar field is present.  This has the flavor of spontaneous symmetry breaking, albeit at zero temperature.  

It would be interesting to understand this zero temperature phase transition from the holographic dual gauge theory perspective, although currently such an interpretation is lacking.  What is known is that in the phase corresponding to the extremal black hole, both holomorphic and anti-holomorphic symmetries are present, whereas in the phase corresponding to the boson star, the anti-holomorphic symmetry is broken by a holomorphic current comprised of a localized scalar field operator.  Whether this is the Goldstone boson corresponding to the spontaneously broken symmetry is currently unknown, although it seems plausible to posit.

Another interesting study would be to consider a massive scalar field, as in \cite{Astefanesei:2003rw}, to see if BTZ black holes emerge as the central energy density of the associated boson star approaches an upper bound; this region of parameter space was unexplored in Ref. \cite{Astefanesei:2003rw}, so it is unclear whether this feature is specific to a massless scalar field or it is more ubiquitous.  What is known is that if a black hole horizon does form around a limiting boson star in the massive scalar field case, it must necessarily also be a degenerate horizon by the mean value theorem.  This can be seen by considering the boundary conditions: at the boson star centre, the metric function $f(r)\approx 1+{\cal O}(r^2)>0$ regardless of the scalar field mass parameter, while asymptotically, $f(r)\approx r^2/\ell^2-M+{\cal O}(r^\alpha)>0$ where $\alpha\le-2$ is a parameter that depends on the scalar field mass.  For a horizon to form around the boson star, $f(r)$ must acquire a root between two boundary points at which $f(r)>0$: the only option is for $f(r)$ to first form a root that is a minimum, \emph{i.e.} a degenerate root, and hence the horizon would be extremal.

Finally, it would be interesting to study the interior solution in the extremal limit to determine its full properties concretely.  As far as the authors are aware, this is the first example of a completely non-singular black hole solution in any dimension that does not resort to string theory or loop quantum gravity inspired models to resolve the singularity.  It would be interesting to construct the interior solution explicitly in order to better understand this singularity-resolved black hole and to see if it can lead to any interesting insights into holography and/or quantum gravity.

\section*{Acknowledgements}

This work was supported by the Natural Sciences and Engineering Research Council of Canada.

\bibliographystyle{plain}

\end{document}